\documentclass[superscriptaddress,floatfix,aps,pre,twocolumn,notitlepage,shortbibliography,nofootinbib]{revtex4-2}

\usepackage[utf8]{inputenc}
\usepackage{amsmath}
\usepackage{amssymb}
\usepackage{refcount}
\usepackage{siunitx}
\usepackage{graphicx}
\usepackage{hyperref}
\hypersetup{colorlinks,allcolors=black}

 \usepackage[capitalise]{cleveref}
\usepackage{microtype}
\usepackage{bm}
\usepackage{lipsum}
\usepackage[english]{babel}
\usepackage{color}
\usepackage{graphicx}
\usepackage{xcolor}
\usepackage{braket}

\begin{document}
\author{Gert Brodin}

\affiliation{Department of Physics, Ume{\aa} University, SE--901 87 Ume{\aa}, Sweden}

\email{gert.brodin@umu.se}
\author{Haidar Al-Naseri}

\affiliation{Stanford PULSE Institute, SLAC National Accelerator Laboratory, Menlo Park, California 94025, USA}
\email{hnaseri@stanford.edu}

\title{Anomalous conductivity due to relativistic Landau quantization}

\begin{abstract} We use a recently developed kinetic model derived from the Dirac equation, in order to study electromagnetic wave propagation in superstrong magnetic fields, such as in magnetars, where relativistic Landau quantization is prominent. The leading contribution to the conductivity tensor in such a plasma is calculated. It is found that the electron Hall current has an anomalous contribution, in the quantum relativistic regime, where the effective particle energy has a significant contribution from the diamagnetic and Zeeman energy. As a result, a new quantum resonance frequency appears, and the dispersion relation for the left- and right-hand polarized modes are strongly modified for long and moderate wavelengths. The implications for magnetar physics are discussed.    
 
\end{abstract}

\maketitle

\section{Introduction}
As is well-known, letting the electromagnetic field strength approach the critical field, opens for a host of new phenomena. For example, aided by a strong electric field, the Schwinger mechanism allows virtual electron-positron pairs to tunnel out of the Dirac sea and become real \cite{fedotov2023advances,di2012extremely}. It may be noted, however, that field strengths of this magnitude cannot yet be produced in the laboratory. In an astrophysical context, on the other hand, magnetic fields of the order of the critical field, and sometimes larger, are known to exist in magnetars \cite{turolla2015magnetars, Strong_magnetic_2020, uzdensky2014plasma}. Also in this case, physics induced by the virtual particles can be crucial, as the vacuum polarization induced by the ultra-strong magnetic fields is believed to be the reason for the observed polarization of electromagnetic radiation emitted from magnetars. Specifically, for photons propagating perpendicular to the magnetic field, the vacuum polarization leads to so-called photon-splitting, implying that photons in one polarization state can decay into photons with a lower frequency belonging to the  opposite polarization state \cite{baring2001photon,turolla2015magnetars}

With real particles present and affected by the magnetic field, either at the magnetar surface or in a magnetar accretion disc, the ultrastrong magnetic field will lead to relativistic Landau quantization \cite{sheng2018wigner,Strong_magnetic_2020,zare2022relativistic,melrose2012spin,brodin2022quantum} of electrons. In a laboratory context, Landau quantization is known to be a key feature in the quantum hall effect of the two-dimensional electron gas \cite{cage2012quantum}, where the transverse resistivity becomes quantized. However, for the case where electrons are subject to relativistic Landau quantization, less is known. While many works have treated certain aspects of relativistic Landau quantization \cite{uzdensky2014plasma,zare2022relativistic,melrose2012spin}, e.g. computed the energy and effective mass of electrons in different Landau states, rigorous treatments of the current response due to an electric field have not been made. On the other hand, when studying electromagnetic wave propagation in magnetar environments, several authors (e.g. \cite{lai2002resonant,lundin2009effective,medvedev2023plasma,wang2007wave,ghosh2002electrical}) have stressed the significance of the strong field vacuum polarization, but at the same time evaluated the current response of the plasma based on classical theory \cite{lai2002resonant,wang2007wave,medvedev2023plasma}, or semi-classical models \cite{lundin2009effective,ghosh2002electrical}, not fully accounting for the dynamics of the perturbed eigenstates.  There is a reasonable justification for this approach since a strong magnetic field mainly influences the perpendicular current response, which tends to be rather effectively suppressed for electrons due to their shorter Larmor radius. Hence there is room for a classical ion response to dominate the perpendicular current. Moreover, the parallel electron current tends to be classical as a rough approximation, since it is only marginally affected by the magnetic field.  While there is a fair bit of merit to this description, nevertheless we will show that the picture given above is an over-simplification

In the present work, we will use a recently developed kinetic approach \cite{Strong_magnetic_2020} in order to evaluate the electron conductivity in a relativistically Landau quantized plasma, of particular relevance for magnetar environments.  We find that the quantum relativistic features induce crucial deviations from the classical conductivity. For certain cases, the idea that relativistically Landau quantized electrons behave almost like classical particles, but with an effective mass dependent on the energy state \cite{uzdensky2014plasma,zare2022relativistic,melrose2012spin}, can be confirmed. However, for other equally common cases, e.g., electromagnetic waves propagating parallel to the magnetic field, the deviations from classical theory turn out to be dramatic. Specifically, both the right and left-hand circular polarized modes behave much differently from classical theory. In this case, one of the wave frequencies approaches zero as the wavenumber $k\rightarrow 0$, and the other approaches a new type of quantum resonance frequency $\omega_{\rm res}$ given by $\omega_{\rm res}\sim (\hbar \omega_c/mc^2) (\hbar \omega_p/mc^2) \omega_p$, where $\hbar$ is the reduced Planck constant, $m$ is the electron mass, $c$ is the speed of light in vacuum, $\omega_c$ is the electron cyclotron frequency, and $\omega_p$ is the electron plasma frequency.  The present results are a prerequisite for understanding the dynamics of relativistically Landau quantized states, as well as for interpreting spectra emitted from magnetars. 
\cite{iqbal2018nonlinear,uzdensky2014plasma,hussain2020influence,hussain2023oblique,areeb2023magnetosonics,jahangir2021nonlinear,eliezer2005effects,sheng2018wigner,medvedev2023plasma}

\section{Basic equations} In this letter, our starting point will be the kinetic
theory derived in Ref. \cite{Strong_magnetic_2020} 
\begin{equation}
\partial _{t}W_{\pm }+\frac{1}{\epsilon }\mathbf{p}\cdot \nabla _{r}W_{\pm
}+q\mathbf{E}\cdot \nabla _{p}W_{\pm }+\frac{1}{\epsilon }\mathbf{p}\times 
\mathbf{B}\cdot \nabla _{p}W_{\pm }  \label{Kinetic_equation}
\end{equation}

This evolution equation is very similar to the usual relativistic Vlasov equation. However, in
order to account for relativistic Landau quantization, the particle energy $%
\epsilon $ is generalized to be a momentum operator  with
\begin{equation}
\epsilon =\sqrt{m^{2}+p^{2}\pm 2\mu _{B}B_{0}-\mu _{B}^{2}B_{0}^{2}(\mathbf{%
\hat{z}}\times \nabla _{p})^{2}}  \label{energy}
\end{equation}%
in units where $c=1$. Here the index $\pm $ on the distribution function
(Wigner function \cite{Wigner-note}) refers to the particle spin state, up or down relative to
the external magnetic field $\mathbf{B}_{0}$,  and $\mu _{B}=e\hbar /2m$ is the Bohr magneton The kinetic evolution equation is combined with Maxwell's equations, where the current and charge densities are given as
\begin{eqnarray}
            {\bf j}=\sum_{\pm}q\int \frac{1}{\epsilon}({\bf p} W_{\pm}) d^3p
\label{current}
\end{eqnarray}
            and
\begin{eqnarray}
          \rho=\sum_{\pm}q\int W_{\pm} d^3p
\end{eqnarray}

The above kinetic theory is based
on a Foldy-Wouthuysen transformation \cite{FoldyWout,Silenko} of the Dirac Hamiltonian,
separating the positive and negative energy states. The model is applicable
for an ultrastrong and homogeneous background magnetic field, $\mathbf{B}%
_{0}=B_{0}\mathbf{\hat{z}}$\thinspace , allowing for magnetic fields $%
B_{0}\sim B_{cr}$ that can be found in magnetars, where the critical field
strength is $B_{cr}=m^{2}c^{3}/|q|\hbar $. Perturbations of the
electromagnetic field added to $\mathbf{B}_{0}$ must still be small compared to the critical field, though,
as the electron and positron states cannot be separated otherwise.  The
model is designed to fully include the effect of relativistic Landau
quantization, but ignore other quantum effects, except degeneracy, that might be included by simply picking a degenerate initial state. Ignoring dynamical quantum effects, besides those due to Landau quantization, is possible if the relevant
dimensionless parameters involving $\hbar $ are small, excluding $\mu
_{B}B_{0}/m\sim 1\,$ which is allowed. To
be concrete, we thus assume $\hbar k\ll p_{th}$(where $p_{th}$ is \
characteristic thermal momentum such that particle dispersive effects can
be ignored), $\hbar k^{2}/m\omega \ll 1$ (making the magnetic dipole force
due to the spin small), together with $\hbar k\ll m$ and $\hbar \omega \ll m$
such that spin-orbit coupling \cite{asenjo2012semi,spin-orbit}and other quantum relativistic effects, except
those due to the strong Landau quantization can be dropped.  In the above, $\omega$ and $k$ represent characteristic frequency and wavenumber scales of the macroscopic variables. 

The physics behind the model is that in a superstrong magnetic field, the
magnetic dipole energy due to the spin and the energy associated with the
orbital angular momentum gives a significant contribution to the particle
energy and thereby to the effective gamma factor $\gamma =\epsilon /m$. This is accompanied by a relativistic momentum spread in the energy eigenstates. In
the expression  (\ref{energy}), the term $\pm 2\mu _{B}B_{0}$ is the energy
contribution from the Zeeman energy due to the spin, whereas  the term $\mu
_{B}^{2}B_{0}^{2}(\mathbf{\hat{z}}\times \nabla _{p})^{2}$ gives the orbital
diamagnetic energy contribution. When the distribution function is in a
Landau quantized energy eigenstate, $W_{\pm }=W_{\pm n}$, we have $\epsilon
W_{\pm n}=$ $\sqrt{m^{2}+p_{z}^{2}+\mu _{B}B_{0}(\pm 1+n)}$ $W_{\pm n}$ (see Eqs. (34)-(35) of Ref.
\cite{Strong_magnetic_2020} for the expression for the energy eigenstate $W_{\pm n}$). However, 
this particular expression is of use mostly to evaluate $\epsilon $ when acting on the
time-independent background, that can be written as a sum of energy
eigenstates. For a perturbation of the background, which in general depends on the azimuthal momentum coordinate, using cylindrical coordinates in momentum space, the distribution function
cannot be written as a sum over energy eigenstates. To evaluate the operator 
$\epsilon $ in this case, we must use the defining expression, where the
root in the energy expression is Taylor expanded to infinite order, 
\begin{equation}
\epsilon =m(1+\frac{1}{2}\frac{%
p^{2}\pm 2\mu _{B}B_{0}-\mu _{B}^{2}B_{0}^{2}(\mathbf{\hat{z}}\times \nabla
_{p})^{2}}{m}+...).
\end{equation}
Naturally, the same definition of $\epsilon $ applies
also when acting on the energy eigenstates, but in this case, the infinite
series sums up to a simple energy eigenvalue.  In all equations written,
note that $\epsilon \ $is acting on all momentum-dependence that
stands to the right, e.g. in the last term of (\ref{Kinetic_equation}), $%
\epsilon $ does not act only on $\nabla _{p}W_{\pm }$, but also on the
momentum in the cross-product $\mathbf{p}\times \mathbf{B}$, and similarly for the energy operator 
$\epsilon$ that appears in the current density (\ref{current}) to be used in Ampere's law.   

\section{Linear theory}
Next, we use \cref{Kinetic_equation} to study linear waves in a magnetized plasma. We divide the variables according to $W=W_0(\mathbf{p})+W_1(\mathbf{r},\mathbf{p},t)$ and $\mathbf{B}=\mathbf{B}_0+\mathbf{B}_1(\mathbf{r},t)$. Furthermore, we use a plane wave ansatz $W_1(\mathbf{r},\mathbf{p},t)=\Tilde{W}_1(\mathbf{p})e^{i(\mathbf{k}\cdot \mathbf{r}-\omega t)}$. Using the ansatz and linearizing \cref{Kinetic_equation}, we obtain
\begin{equation}
  \Big[-i\omega +\frac{\mathbf{k}}{\epsilon}\cdot \mathbf{p} + \frac{q B_0}{\epsilon}\Big]\Tilde{W}_1= -q\Big[\mathbf{E}+  \frac{1}{\epsilon} \mathbf{p}\times \mathbf{B}_1\Big]\cdot \nabla_pW_0
\end{equation}
Due to the energy $\epsilon$ being a momentum operator, the standard techniques that work for the linearized Vlasov equation are not directly applicable. However, the equation can be solved using an expansion of $\Tilde{W}_1$ according to
\begin{equation}
    \Tilde{W}_1=\sum_{n=-\infty}^{\infty}\sum_{m=0}^{\infty} g_{nm}(p_{\bot},p_z)e^{in\varphi_p} \Big(\frac{k_{\bot}p_{\bot}}{m\omega_c} \Big) \label{expansion}
\end{equation}
In principle, we can include terms to an arbitrary order in the summation over $n$ and $m$. However, in the strongly magnetized regime that we consider (where all other frequencies are small compared to the electron cyclotron frequency), only the low values of $n$ and $m$ will give a significant contribution. Moreover, without loss of generality we let the wavevector be directed in the $xy$-plane, i.e. ${\bf k}=k_x{\hat {\bf x}}+k_z{\hat {\bf z}}$, such that $k_{\perp}=k_x$. 
Using the expansion (\ref{expansion}), we solve 
for $\Tilde{W}_1$ to leading order, including terms up to $n=\pm 1$ and $m=1$.
Combining the solution for $\Tilde{W}_1$ with Ampere's law, we obtain the dispersion relation for electromagnetic wave propagation (including also
the electrostatic limiting cases) in a magnetized plasma, extending previous classical results \cite{stix1992waves} to account for relativistic Landau quantization. The dispersion relation can be written as $\det
D_{ij}=0$, where the matrix is written as%

\begin{equation}
    D_{ij}=\delta _{ij}\bigg(1-\frac{k^{2}c^{2}}{\omega ^{2}}\bigg)+\frac{k_{i}k_{j}c^{2}}{%
\omega ^{2}}+\chi _{ij}
\end{equation}
where $\delta _{ij}$ comes from the displacement current in Ampere's law,
and the terms proportional to the wavenumber comes from the curl of the
magnetic field. The last term, $\chi _{ij}\,,$ is the plasma susceptibility
and is related to the plasma currents as $\chi _{ij}=\sigma _{ij}/i\omega
\varepsilon _{0}$, where $\sigma _{ij}$ is the conductivity tensor
determined from the kinetic evolution equation. With the wavevector in the xz-plane, the expression for $%
D_{ij}$ becomes: 
\begin{equation}
D_{ij}=\left[ 
\begin{array}{ccc}
1-\frac{k_{z}^{2}c^{2}}{\omega ^{2}}+\chi _{xx} & \chi _{yx} & \frac{%
k_{x}k_{z}c^{2}}{\omega ^{2}}+\chi _{zx} \\ 
\chi _{yx}^{\ast } & 1-\frac{k^{2}c^{2}}{\omega ^{2}}+\chi _{yy} & \chi _{yz}
\\ 
\frac{k_{x}k_{z}c^{2}}{\omega ^{2}}+\chi _{zx}^{\ast } & \chi _{yz}^{\ast }
& 1-\frac{k_{x}^{2}c^{2}}{\omega ^{2}}+\chi _{zz}%
\end{array}%
\right] 
\end{equation}

where 

\begin{widetext}
    
\begin{align}
\chi_{xy}=\chi_{yx}^{*}&=-\frac{i\omega q 2\pi }{4} \int d^2p\frac{1}{\epsilon} \frac{\pm p_{\bot}}{\omega-kp_z/\epsilon' \mp qB_0/\epsilon'}
    \bigg[ 
    \frac{\partial f_0}{\partial p_{\bot}}
    -\frac{k_z}{\omega} \frac{1}{\epsilon'}
     \Big( p_z\frac{\partial W_0}{\partial p_{\bot}}-p_{\bot}\frac{\partial W_0}{\partial p_z} \Big)
    \bigg]\\
    \chi_{yz}=\chi_{zy}^{*}&=\frac{i\omega q 2\pi }{4} \int d^2p\frac{1}{\epsilon} \frac{\mp p_{\bot}}{\omega-kp_z/\epsilon' \mp qB_0/\epsilon'}
    \bigg[ 
    \frac{k_z}{\omega} \frac{1}{\epsilon'}
     \Big( p_z\frac{\partial W_0}{\partial p_{\bot}}-p_{\bot}\frac{\partial W_0}{\partial p_z} \Big)
    \bigg]
\end{align}

\end{widetext}
The given components for the electron susceptibility constitute the dominating contributions for electrons in the regime $\omega\ll \omega_c=qB_0/m\sim qB_0/\epsilon'$. Accordingly, we should replace the denominators in the expressions above according to $\omega-kp_z/\epsilon' \mp qB_0/\epsilon'\rightarrow qB_0/\epsilon'$. For consistency with other approximations, this should be done whenever applying the expressions. The only reason to show the formula without this simplification is to emphasize the similarity of the quantum result with previous well-known classical formulas, when the arguments of the Bessel functions are small \cite{stix1992waves}. The expression for $\chi _{zz}$ is not given above, but we note that this term has been computed previously, see Ref. \cite{Strong_magnetic_2020}. Contrary to the terms given here, the component for $\chi _{zz}$ can essentially be written as for a classical plasma, except that each Landau-quantized energy eigenstate behaves like a separate particle species, which contributes to the plasma frequency with its own effective mass, see Ref. \cite{Strong_magnetic_2020} for details. 

The remaining susceptibility components for electrons are approximated with zero, since in the regime $\omega\ll \omega_c$ studied here, those components will always be negligible compared to the ion contribution. While the ion contribution has not been written out, a contribution from the ions can always be added to give the total susceptibility. The ions susceptibility will typically be given by the classical textbook results (see e.g. Ref. \cite{stix1992waves}), since even for magnetar field strengths $\hbar \omega_{ci}\ll k_B T$, where $\omega_{ci}$ is the ion cyklotron frequency, $k_B$ is the Boltzmann constant and $T$ is the temperature (the condition for neglecting ion Landau quantization) tend to apply.         

\section{The linear Dispersion relation}

Let us next consider the dispersion relation for electromagnetic waves
propagating parallel to the external magnetic field.  Although we are interested in the case of ultra-strong magnetic fields, approaching the critical field, in order to obtain simple analytical results, we will make an expansion in the parameter $\mu
_{B}B_{0}/m\ll 1$. For $\omega /\omega
_{ce}\ll 1\,$, we keep terms  to first order in $\omega /\omega _{ce}$,
but allow for $\omega /\omega _{ci}\sim 1$. Moreover, we keep correction terms to second order in $\mu_{B}B_{0}/m$. 
By considering an electron-ion plasma, and
evaluating the ion current in the classical cold limit, we find that the total (including electrons and ions) susceptibility is given by
$\chi _{yx}=\chi _{yx,e}+\chi _{yx,i}$
with the electron contribution
\begin{equation}
\chi _{yx,e} =\frac{\omega _{pe}^{2}}{\omega _{ce}\omega }\left[ 1+\frac{3}{2}
\left( \frac{\mu _{B}B_{0}}{m}\right) ^{2}\right]
\end{equation}
and the ion contribution
\begin{equation}
\chi _{yx,i}=\frac{\omega
_{pi}^{2}\omega _{ci}}{\omega _{ci}^{2}-\omega ^{2}}
\end{equation}
such that the total susceptibility is
\begin{eqnarray}
\chi _{yx}
&=&\frac{\omega _{pe}^{2}}{\omega _{ce}\omega }\left[ \frac{\omega ^{2}}{%
\omega ^{2}-\omega _{ci}^{2}}+\frac{3}{2} \left( \frac{\mu _{B}B_{0}}{m}\right) ^{2}%
\right]  \label{EQ10}
\end{eqnarray}
Consequently, for reasonably low frequencies, $\omega \ll \omega _{ci},$ the second term $\propto (\mu_B B_0/m)^2$,
which is the quantum contribution associated with the electron Landau quantization, will dominate for magnetic fields not too much smaller than the critical
field. Calculating the determinant of the dispersion matrix $D_{ij}$ for $%
k_{\bot }=0$, we obtain the dispersion relation   
\[
\left( 1-\frac{k^{2}c^{2}}{\omega ^{2}}-\frac{\omega _{pi}^{2}}{\omega
^{2}-\omega _{ci}^{2}}\right) =
\pm \frac{3}{2} \frac{\omega _{pe}^{2}}{\omega _{ce}\omega }%
\left( \frac{\mu _{B}B_{0}}{m}\right) ^{2}  
\]%
where the two signs correspond to the left and right hand circularly
polarized modes, respectively. Naturally, the dispersion relation can be
studied in all generality. However, to illustrate the main new features we
limit ourselves to frequencies well below the ion-cyclotron frequency, in
which case we obtain
\begin{equation}
\left( 1-\frac{k^{2}c^{2}}{\omega ^{2}}+\frac{\omega _{pi}^{2}}{\omega
_{ci}^{2}}\right) =\pm \frac{3}{2} \frac{\omega _{pe}^{2}}{\omega _{ce}\omega }%
\left( \frac{\mu _{B}B_{0}}{m}\right) ^{2}  \label{DR1}
\end{equation}
In the short-wavelength limit, the quantum term becomes negligible, and we get 
\begin{equation}
\omega ^{2}=\frac{k^{2}c^{2}}{1+\omega _{pi}^{2}/\omega _{ci}^{2}}
\end{equation}
for both the left and right hand circularly polarized mode. For $\omega $
approaching $\omega _{ci}$, naturally we should avoid the low-frequency
simplification and the left and right hand modes will have different frequencies also in the classical limit. The
solution to \ref{DR1}  is 

\begin{figure}
    \centering
    \includegraphics[width=9 cm, height= 8 cm]{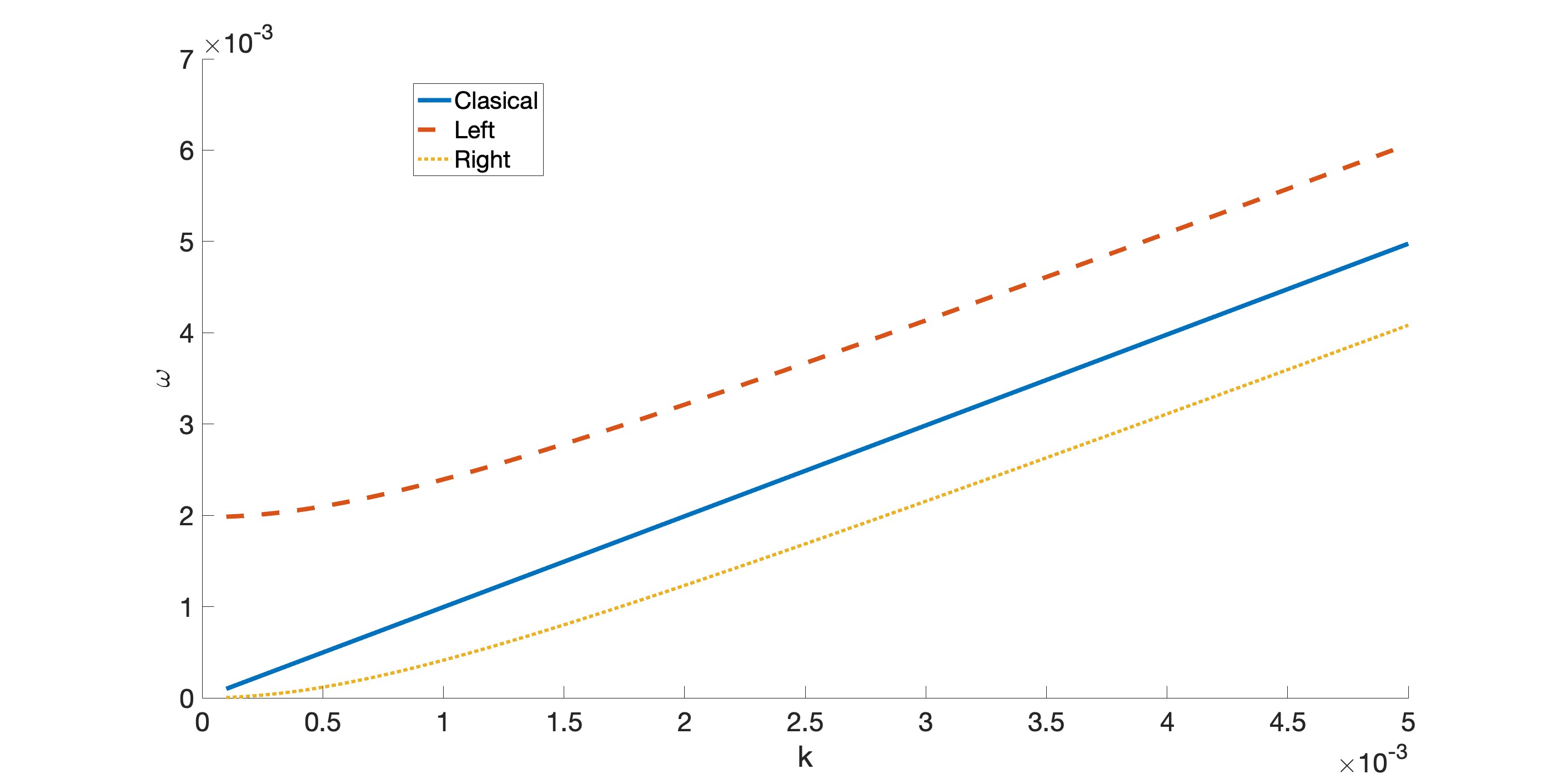}
    \caption{The frequency $\omega$ is plotted as a function of the wave vector $k$ for $\mu_BB_0/m =0.2$ and $\omega_c/\omega_p=10$.}
    \label{FIG1}
\end{figure}

\begin{eqnarray}
\omega =-\frac{3 \omega _{pe}^{2}}{4\omega _{ce}\left( 1+\omega
_{pi}^{2}/\omega _{ci}^{2}\right) }\left( \frac{\mu _{B}B_{0}}{m}\right)
^{2} \nonumber
\\ 
\pm \sqrt{\left[ \frac{3 \omega _{pe}^{2}}{4\omega _{ce}\left(
1+\frac{\omega _{pi}^{2}}{\omega _{ci}^{2}}\right) }\left( \frac{\mu _{B}B_{0}}{m}%
\right) ^{2}\right] ^{2}+\frac{k^{2}c^{2}}{\left( 1+\frac{\omega _{pi}^{2}}{\omega _{ci}^{2}}\right) }} \label{analytical}
\end{eqnarray}
Thus in the long wavelength limit, we have a new type of quantum resonance
frequency $\omega _{q}$, such that when $k\rightarrow 0$, for the left
hand circularly polarized mode we get $\omega \rightarrow \omega
_{q}$ with 
\begin{equation}
\omega _{q}=-\frac{3 \omega _{pe}^{2}}{2\omega _{ce}\left( 1+\omega
_{pi}^{2}/\omega _{ci}^{2}\right) }\left( \frac{\mu _{B}B_{0}}{m}\right) ^{2} \label{qres}
\end{equation}
whereas the right-hand circularly polarized mode fulfills  
\begin{equation}
\omega \simeq \frac{k^{2}c^{2}}{4\omega _{q}\left( 1+\omega _{pi}^{2}/\omega
_{ci}^{2}\right) }
\end{equation}
for long wavelengths fulfilling $\frac{k^{2}c^{2}}{\left( 1+\omega
_{pi}^{2}/\omega _{ci}^{2}\right) }\ll \omega _{q}^{2}$. We thus see that
for a sufficiently strong magnetization of the plasma, when relativistic
Landau quantization is significant, both the left and right- hand circularly
polarized modes show a distinct quantum behavior for long wavelengths, see \cref{FIG1} where the left- and right-handed solutions to \cref{DR1} are compared with the classical limit. For shorter wavelengths, $%
\frac{k^{2}c^{2}}{\left( 1+\omega _{pi}^{2}/\omega _{ci}^{2}\right) }\gg
\omega _{q}^{2}$ not covered in \cref{FIG1} , the quantum behavior is suppressed, and we recover the
classical limit. The wavelength where the transition from quantum to
classical behavior takes place depends on the plasma density. For the high
densities at the surface of magnetars, the plasma frequency is of the order $%
\omega _{pe}\sim 10^{19-20}s^{-1}$. In this case, for magnetars with surface
magnetic fields slightly below the critical field, the quantum limits of the dispersion relation shown above apply
for wavelengths in the UV region and longer.


\section{Discussion and conclusion}
As is known from condensed matter physics, and seen in the quantum Hall effect, the effect of Landau quantization may induce peculiar properties of the conductivity. In condensed matter systems, this may happen already in a non-relativistic description. In that case, besides Landau quantization, the special quantum features are dependent on the system being a 2D electron gas. In the present theory, the electrons are not limited to two dimensions, but instead, the relativistic aspects of Landau quantization are crucial, accounting for the contribution to the electron's effective mass through the diamagnetic and Zeeman energies. At first glance, the extra quantum contributions to the Hall term of the electron susceptibility (leading to the quantum contribution in the total susceptibility (\ref{EQ10}), i.e., the second term $\propto (\mu_B B_0/m)^2$) may be expected. Why would the susceptibility (conductivity) be unaffected when the Zeeman energy becomes significant? However, looking more carefully at the result, it appears to challenge a key principle of physics, namely Lorentz invariance. 

Normally, in the limit when $k\rightarrow 0$ and $\omega \rightarrow 0$, the total current density of all species cancels.
The reason is as follows: For an electromagnetic field that is effectively
static and homogenous, in a reference system moving with the $\mathbf{%
E\times B}-$drift, where $\mathbf{v}_{E}=\mathbf{E\times B/}B^{2}$,
particles will only feel a static magnetic field. Hence, independent of
species, there will only be a gyration and no net-drift in this system. This
corresponds to all species $s$ having the current density $q_{s}n_{0s}%
\mathbf{v}_{E}$ in the original system. Due to the background neutrality of
all species, this implies a vanishing total current density. Fundamentally,
a vanishing current density in the limit $k\rightarrow 0$ and $\omega
\rightarrow 0$ is a classical result that follows from Lorentz invariance.
In the present case, while our governing equations do not show Lorentz
invariance directly (since they are built around a preferred reference
system subject to a strong magnetic field $B_{0}$), we should still demand
our results to be consistent with Lorentz invariance. Clearly, since the
model is derived from the Dirac equation and Maxwell's equations, we should
not have a theory that is in conflict with the basic principles of special
relativity. Thus the question arises, how can the effective electron drift
velocity deviate from the $\mathbf{E\times B}-$drift in the limit $%
k\rightarrow 0$ and $\omega \rightarrow 0$, without contradicting Lorentz
invariance? 

A first observation is that, contrary to classical theory, it is no longer
meaningful to think of individual velocities for different phase space
elements, represented by $\mathbf{p}/\varepsilon $. In the limit where this
would be accurate, the quantum contribution to $\chi _{xy}$ vanish, and
the current density for electrons indeed become $q_{e}n_{0e}\mathbf{v}_{E}$.
A complication of the present theory, where the energy $\varepsilon $ is a
momentum-operator, is that the effective velocity of the theory depends on
the behavior in a region of momentum space, rather than in a single point. 
Moreover, the effective velocity averaged over all momentum for energy
transport ($\mathbf{v}_{en}=\int \frac{1}{\varepsilon }(\mathbf{p}%
\varepsilon f)d^{3}p/\int \varepsilon fd^{3}p$)  and for particle transport (%
$\mathbf{v}_{p}=\int \frac{1}{\varepsilon }(\mathbf{p}f)d^{3}p/\int f$ ) 
only coincide if the distribution function $f$ is an energy eigenstate, in
general $\mathbf{v}_{en}\neq $  $\mathbf{v}_{p}$. However, even when this
ambiguity regarding what constitutes the velocity is noted, one can argue
that in a reference system with a static magnetic field, there should be no
net current density. If this were true, one would obtain the same
cancellation of the total current density as in the classical case. 

This argument leaves out a key aspect of the present theory, though. When a
Landau quantized eigenstate in a pure magnetic field is exposed to an
electric field, for the particle to move at all, the momentum distribution
of the particles need to be modified in a way that prevents the particle state from being written as a sum of energy eigenstates (this is not possible, as we break the angular symmetry and get a dependence on $\varphi _{p}$). Breaking
the original symmetry contributes to the energy by a positive and quantized
amount, since for the perturbed distribution function we must let $%
B_{0}^{2}\left( \frac{\partial ^{2}}{\partial \rho ^{2}}+\frac{1}{\rho }%
\frac{\partial }{\partial \rho }\right) \rightarrow B_{0}^{2}\left( \frac{%
\partial ^{2}}{\partial \rho ^{2}}+\frac{1}{\rho }\frac{\partial }{\partial
\rho }+\frac{1}{\rho ^{2}}\right) $ in the orbital magnetic energy part of
the energy operator. This quantized change of energy when modifying the
particle state has no classical correspondence and is what determines the
momentum-velocity relation when calculating the magnitude of the perturbed
distribution function. Importantly, the effective velocity obtained in this
step is not the same as the effective velocity that we get when calculating
the current density. Here, for the part surviving the momentum integration,
the energy operator acts on a function of momentum that is angular
symmetric, which leads to a different expression of what can be interpreted
as the effective velocity. Ultimately, in the relativistically
Landau quantized regime, what constitutes "particle velocity" depends on the details of the definition. As a result, a ratio of different particle energies, which would simply be unity in a classical calculation, is what induces a relativistically quantum-corrected Hall current.  

The main conclusion from our analysis is that the anomalous conductivity for relativistically Landau quantized states can modify the electromagnetic wave propagation properties in magnetar environments in a distinct way. However, it is a challenge to relate the properties of the plasma susceptibility tensor to observational magnetar data. Nevertheless, a specific possibility might be to look for the quantum resonance given by \cref{qres} in the observed spectra.

In the present work, we have focused on the simplest aspects of the linearized theory of electromagnetic waves. A more complete study of the linear susceptibility and generalizations to cover also the nonlinear regime are projects for future work. It may then be of interest to include also the physics associated with the magnetic dipole force due to the electron spin \cite{brodin2022quantum,asenjo2012semi}, and the vacuum polarization associated with strong magnetic fields, see e.g. Refs. \cite{haas2010fluid,lundin2009effective,eriksson2004possibility,uzdensky2014plasma} 

\section{acknowledgement}
We are grateful to Frederico Fiuza and Christopher Thompson for valuable discussions. H. A-N also like to acknowledge support
by the Knut and Alice Wallenberg Foundation through the fellowship KAW 2022.0361.

 \bibliography{References}

\begin{thebibliography}{30}
\bibitem [{Wig()}]{Wigner-note}%
  \BibitemOpen
  \href@noop {} {\bibinfo  {journal} {Formally, since the derivation uses a Wigner-transform, the function $W$ is a Wigner-function, rather than a distribution function. As a result, it can also be negative in limited regions of momentum space. However, due to the similarity with the Vlasov equation, for most practical purposes, it makes sense to think of $W$ as a distribution function, and we will therefore refer to it as such.}\ }\BibitemShut {NoStop}%
\end{thebibliography}


\end{document}